\documentclass[Physsubmission, Phys]{SciPost}

\newcommand{\Pom}{\mathbb{P}}

\newcommand{\Reg}{\mathbb{R}}

\newcommand{\bpta}{\mbox{\boldmath $p_{t,1}$}}
\newcommand{\bptb}{\mbox{\boldmath $p_{t,2}$}}
\newcommand{\bkt}{\mbox{\boldmath $k_{t}$}}
\newcommand{\bktsqrt}{\mbox{\boldmath{$k_{t}^{2}$}}}
\newcommand{\bptat}{\mbox{\boldmath $\tilde{p}_{t,1}$}}
\newcommand{\bptbt}{\mbox{\boldmath $\tilde{p}_{t,2}$}}
\newcommand{\p}{\partial}
\newcommand{\twosidep}[1]{\stackrel{\leftrightarrow}{\p}_{\! #1}}

\newcommand*\wideestimates{\mathrel{\widehat{=}}}

\hypersetup{
    colorlinks,
    linkcolor={red!50!black},
    citecolor={blue!50!black},
    urlcolor={blue!80!black}
}

\usepackage[bitstream-charter]{mathdesign}
\DeclareSymbolFont{usualmathcal}{OMS}{cmsy}{m}{n}
\DeclareSymbolFontAlphabet{\mathcal}{usualmathcal}

\begin{document}

\begin{center}{\Large \textbf{
Central exclusive production of axial-vector $f_{1}$ mesons
in proton-proton collisions
within the tensor-pomeron approach\\
}}\end{center}

\begin{center}
Piotr Lebiedowicz\textsuperscript{$\star$}
\end{center}

\begin{center}
Institute of Nuclear Physics Polish Academy of Sciences,\\ Radzikowskiego 152, PL-31342 Krak{\'o}w, Poland
\\
* Piotr.Lebiedowicz@ifj.edu.pl
\end{center}

\begin{center}
\today
\end{center}


\definecolor{palegray}{gray}{0.95}
\begin{center}
\colorbox{palegray}{
  \begin{minipage}{0.95\textwidth}
    \begin{center}
    {\it  XXXIII International (ONLINE) Workshop on High Energy Physics \\“Hard Problems of Hadron Physics:  Non-Perturbative QCD \& Related Quests”}\\
    {\it November 8-12, 2021} \\
    \doi{10.21468/SciPostPhysProc.?}\\
    \end{center}
  \end{minipage}
}
\end{center}

\section*{Abstract}
{\bf
We discuss the central exclusive production of $f_{1}$ mesons 
in proton-proton collisions. 
The diffractive pomeron-pomeron fusion process 
within the tensor-pomeron approach is considered.
Two ways to construct the pomeron-pomeron-$f_{1}$ coupling are discussed.  
The theoretical calculation of coupling constants 
is a challenging problem of nonperturbative QCD. 
We adjust the parameters of the model to the WA102 experimental data.
The total cross section and differential distributions are presented. 
Predictions for LHC experiments are given.
Detailed analysis of the distributions in $\phi_{pp}$
the azimuthal angle between the transverse momenta of the outgoing protons can help to check different models and to study real pattern of the absorption effects.}

\vspace{10pt}
\noindent\rule{\textwidth}{1pt}
\tableofcontents\thispagestyle{fancy}
\noindent\rule{\textwidth}{1pt}
\vspace{10pt}

\section{Introduction}
\label{sec:intro}

\begin{figure}[h]
\centering
\includegraphics[width=.3\textwidth]{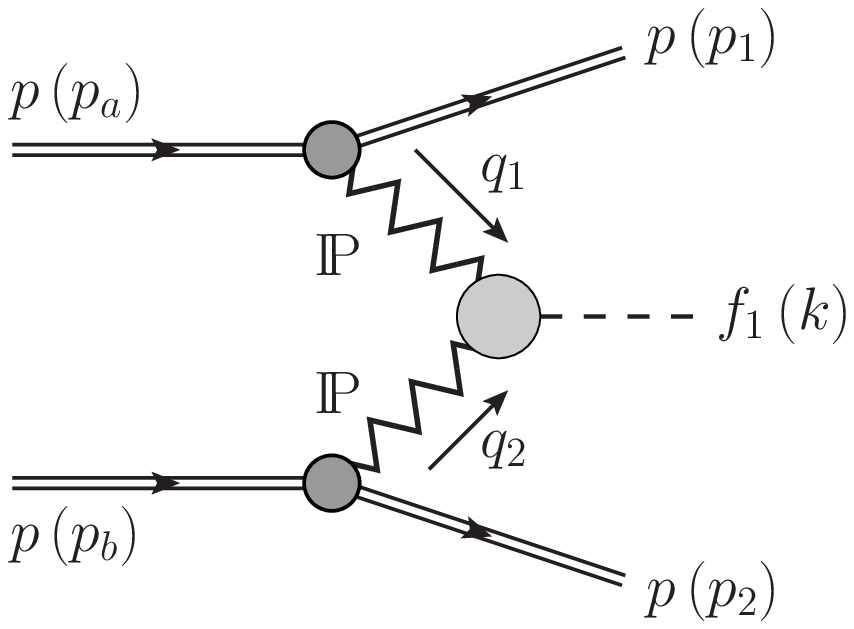}\qquad
\includegraphics[width=.3\textwidth]{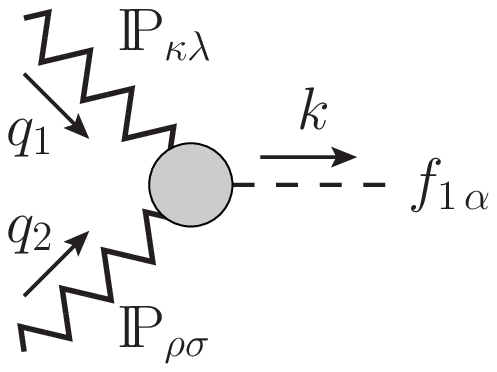}
\caption{Diagrams for the reaction (\ref{1.1}) 
with double-$\Pom$ exchange and the $\Pom \Pom f_{1}$ vertex.}
\label{fig1}
\end{figure}

In this contribution we discuss central exclusive
production (CEP) of axial-vector $f_{1}$ 
($J^{PC} = 1^{++}$) mesons
in proton-proton collisions
\begin{eqnarray}
p(p_{a},\lambda_{a}) + p(p_{b},\lambda_{b}) \to
p(p_{1},\lambda_{1}) + f_{1}(k ,\lambda) + p(p_{2},\lambda_{2}) \,,
\label{1.1}
\end{eqnarray}
where $p_{a,b}$, $p_{1,2}$ and $\lambda_{a,b}$, 
$\lambda_{1,2} = \pm 1/2$
denote the four-momenta and helicities of the protons, 
and $k$ and $\lambda = 0, \pm 1$ 
denote the four-momentum and helicity of the $f_{1}$ meson, respectively.
Here $f_{1}$ stands for one of the axial-vector mesons 
with $J^{PC} = 1^{++}$,
i.e. $f_{1}(1285)$ or $f_{1}(1420)$.
This presentation summarises some of the
key results of \cite{Lebiedowicz:2020yre} 
to which we refer the reader for further details.
CEP of $f_{1}(1285)$ and $f_{1}(1420)$ mesons
was measured by WA102 Collaboration
\cite{Barberis:1998by,Barberis:1999wn,Kirk:1999df}.
Their internal structure ($q \bar{q}$, tetraquark, $K\bar{K}$ molecule) 
remains to be established.
At high energies the double-pomeron exchange
mechanism (Figure~\ref{fig1}) is expected to be dominant.
The pomeron ($\Pom$) is essential object for understanding
diffractive phenomena. 
Within QCD is a color singlet, predominantly gluonic object, 
thus the CEP of mesons has long been regarded as a potential
source of glueballs.

For soft reactions,
calculations of the pomeron from first principle 
are currently not possible, and one has to retreat
to Regge models to describe soft high-energy diffractive 
scattering.
Until recently, the spin structure of the pomeron has not 
received much attention. 
It is well known that the pomeron carries vacuum quantum numbers 
with regard to charge, color, isospin and charge conjugation. 
But what about spin? 
It has been shown some time ago that 
the charge-conjugation $C = +1$ pomeron can
be regarded as a coherent sum of elementary spin $2 + 4 + 6 + \ldots$
exchanges \cite{Nachtmann:1991ua}.
The tensor-pomeron model introduced in \cite{Ewerz:2013kda}
assumes this property.
We treat the reaction (\ref{1.1}) in this model,
in which the pomeron exchange is described
as effective rank 2 symmetric tensor exchange.
This approach has a good basis from nonperturbative QCD
using functional integral techniques \cite{Nachtmann:1991ua}.
A tensor character of the pomeron is also preferred
in holographic QCD models
\cite{Brower:2006ea,Domokos:2009hm,Ballon-Bayona:2015wra,Iatrakis:2016rvj}.

The tensor-pomeron model 
was applied to two-body hadronic reactions 
\cite{Ewerz:2013kda,Ewerz:2016onn,Lebiedowicz:2021byo},
to photoproduction of $\pi^{+} \pi^{-}$ pairs \cite{Bolz:2014mya},
to low-x deep inelastic lepton-nucleon scattering 
and photoproduction \cite{Britzger:2019lvc}, 
and especially to CEP reactions
\begin{eqnarray}
&&p + p \to p + X + p\,, \nonumber\\
&&X = \eta,\, \eta',\, f_{0},\, f_{1},\, f_{2},\, \pi^{+}\pi^{-},\,
K^{+}K^{-},\, p \bar{p},\, 4\pi,\, 4K,\, \rho^{0},\, \phi,\, \phi \phi,\, K^{*0} \bar{K}^{*0}\,;
\end{eqnarray}
see e.g. \cite{Lebiedowicz:2013ika,Lebiedowicz:2016ioh,Lebiedowicz:2016zka,Lebiedowicz:2019por,Lebiedowicz:2019boz,Lebiedowicz:2019jru,Lebiedowicz:2021pzd}.
In this model the $C = -1$ odderon \cite{TOTEM:2020zzr}
is described as effective vector exchange.
Exclusive reactions suitable for studies of the odderon exchange
at high energies were discussed in
\cite{Bolz:2014mya,Lebiedowicz:2019boz,Lebiedowicz:2019jru}.
Conceptually, vector-type couplings of the pomeron turn out to be rather questionable. 
For example, a vector pomeron implies that the total cross sections 
for $pp$ and $p\bar{p}$ scattering at high energy have opposite sign
\cite{Ewerz:2016onn}. 
But, of course, quantum field theory forbids negative cross sections. 
A further argument against a vector pomeron was shown in \cite{Britzger:2019lvc},
mainly it does not give any contribution to photoproduction data. 
One may also ask about the possibility of a scalar coupling of the
pomeron to external particles. 
While possible from the point of view of QFT, 
such a coupling is experimentally disfavoured. 
In \cite{Ewerz:2016onn} it was shown that STAR data \cite{Adamczyk:2012kn}
on polarised elastic $pp$ scattering 
are compatible with the tensor-pomeron ansatz
but clearly rule out a scalar character of the soft pomeron
what its coupling concerns.
Also some historical remarks on different views of the pomeron 
were made in \cite{Ewerz:2016onn}.
In the light of our discussion here we cannot support
the conclusions of \cite{Close:1999is,Close:1999bi}
that the pomeron behaves (couples) like vector current.

The theoretical calculation of $\Pom \Pom f_{1}$ coupling
is a challenging problem of nonperturbative QCD.
We argue that the pomeron couplings play
an important role, and that they should be treated as tensor couplings.
Using our model we perform a fit to the available WA102 data
\cite{Barberis:1998by,Kirk:1999df}
and we analyse whether our study could shed light 
on the $\Pom \Pom f_{1}$ couplings.
In the future the model parameters 
($\Pom \Pom f_{1}$ coupling constants, 
cutoff parameters in form factors)
could be adjusted by comparison with precise experimental data 
from both RHIC and the LHC.
The $\pi^{+}\pi^{-}\pi^{+}\pi^{-}$ channel seems 
well suited to measure the $f_{1}(1285)$ CEP at high energies.
For a preliminary data of the reaction
$pp \to pp 2\pi^{+}2\pi^{-}$
measured at LHC@13TeV by the ATLAS Collaboration
see \cite{Sikora:2020mae}.

\section{Sketch of the formalism}
\label{sec:formalism}

\subsection{The amplitude for the $pp \to pp f_{1}$ reaction}

The Born-level amplitude for the reaction (\ref{1.1})
via pomeron-pomeron fusion (Figure~\ref{fig1})
can be written as
\begin{eqnarray}
{\cal M}^{\rm Born}_{\lambda_{a} \lambda_{b} \to \lambda_{1} \lambda_{2} \lambda}
&=& (-i)\, (\epsilon^{\mu}(\lambda))^{*}\,
\bar{u}(p_{1}, \lambda_{1}) 
i\Gamma^{(\Pom pp)}_{\mu_{1} \nu_{1}}(p_{1},p_{a}) 
u(p_{a}, \lambda_{a}) \nonumber \\
&& \times 
i\Delta^{(\Pom)\, \mu_{1} \nu_{1}, \alpha_{1} \beta_{1}}(s_{1},t_{1}) \,
i\Gamma^{(\Pom \Pom f_{1})}_{\alpha_{1} \beta_{1}, \alpha_{2} \beta_{2}, \mu}(q_{1},q_{2}) \,
i\Delta^{(\Pom)\, \alpha_{2} \beta_{2}, \mu_{2} \nu_{2}}(s_{2},t_{2}) \nonumber \\
&& \times 
\bar{u}(p_{2}, \lambda_{2}) 
i\Gamma^{(\Pom pp)}_{\mu_{2} \nu_{2}}(p_{2},p_{b}) 
u(p_{b}, \lambda_{b}) \,.
\label{amplitude_f1_pompom}
\end{eqnarray}
The relevant kinematic quantities are
\begin{eqnarray}
&&s = (p_{a} + p_{b})^{2},\;
s_{1} = (p_{a} + q_{2})^{2} = (p_{1} + k)^{2},\;
s_{2} = (p_{b} + q_{1})^{2} = (p_{2} + k)^{2},
\nonumber\\
&&
k = q_{1} + q_{2},\;
q_1 = p_{a} - p_{1}, \;q_2 = p_{b} - p_{2},\;
t_1 = q_{1}^{2}, \;
t_2 = q_{2}^{2}, \;
m_{f_{1}}^{2} = k^{2}\,,
\label{1.2}
\end{eqnarray}
In (\ref{amplitude_f1_pompom}) 
$\epsilon^{\mu}(\lambda)$ is the polarisation vector of the $f_{1}$ meson,
$\Delta^{(\Pom)}$ and $\Gamma^{(\Pom pp)}$ 
denote the effective propagator and proton vertex function, respectively, 
for the tensor-pomeron exchange \cite{Ewerz:2013kda}.
The new quantity, to be studied here, is the $\Pom \Pom f_{1}$ coupling
(vertex function).
In our analysis we should also 
include absorption effects to the Born amplitude.
Then the full amplitude is
\begin{eqnarray}
{\cal {M}}_{pp \to pp f_{1}} =
{\cal {M}}_{pp \to pp f_{1}}^{\rm Born} + 
{\cal {M}}_{pp \to pp f_{1}}^{pp-\rm{rescattering}}\,.
\label{amp_full}
\end{eqnarray}
The amplitude including the $pp$-rescattering corrections
can be written as (within the one-channel-eikonal approach)
\begin{eqnarray}
{\cal M}_{pp \to pp f_{1}}^{pp-\rm{rescattering}}(s,\bpta,\bptb)=
\frac{i}{8 \pi^{2} s} \int d^{2}\bkt \,
{\cal M}_{pp\to pp f_{1}}^{\rm Born}(s,\bptat,\bptbt)
{\cal M}_{pp \to pp}(s,t)\,,
\label{abs_correction}
\end{eqnarray}
where $\bpta$ and $\bptb$
are the transverse components of the momenta of the outgoing protons
and $\bkt$ is the transverse momentum carried around the pomeron loop.
${\cal M}_{pp\to pp f_{1}}^{\rm Born}$
is the Born amplitude given by (\ref{amplitude_f1_pompom})
with $\bptat = \bpta - \bkt$ and $\bptbt = \bptb + \bkt$.
${\cal M}_{pp \to pp}$
is the elastic $pp$ scattering amplitude
for large $s$ and with the momentum transfer $t=-\bktsqrt$.
In practice we work with the amplitudes in the high-energy approximation,
i.e. assuming $s$-channel helicity conservation 
in the pomeron-proton vertex.

\subsection{The pomeron-pomeron-$f_{1}$ coupling}




We follow two strategies for constructing 
the $\Pom \Pom f_{1}$ coupling and the vertex
function.

\textbf{(1)} Phenomenological approach.
First we consider a fictitious process: the fusion of two
``real spin-2 pomerons'' (or tensor glueballs) of mass $m$
giving an $f_{1}$ meson of $J^{PC} = 1^{++}$.
We make an angular momentum analysis of this reaction 
in its c.m. system, the rest system of the $f_{1}$ meson:
$\Pom\,(m, \epsilon_{1}) + \Pom\,(m, \epsilon_{2}) \to
f_{1}\,(m_{f_{1}},\epsilon)$.
The spin~2 of these ``pomerons'' can be combined to a total spin $S$
($0 \leqslant S \leqslant 4$) and this must be combined with
the orbital angular momentum $l$ to give the $J^{PC} = 1^{++}$
values of the $f_{1}$.
There are two possibilities,
$(l,S) = (2,2)\; \rm{and}\; (4,4)$
(see Appendix~A of \cite{Lebiedowicz:2013ika}),
and corresponding coupling Lagrangians $\Pom \Pom f_{1}$ are:
\begin{eqnarray}
&&
{\cal L}^{(2,2)}_{\Pom \Pom f_{1}} = \frac{g'_{\Pom \Pom f_{1}}}{32\,M_{0}^{2}}
\Big( \Pom_{\kappa \lambda} 
\twosidep{\mu} \twosidep{\nu}
\Pom_{\rho \sigma} \Big)
\Big( \p_{\alpha} U_{\beta} - \p_{\beta} U_{\alpha} \Big)\,
\Gamma^{(8)\,\kappa \lambda, \rho \sigma, \mu \nu, \alpha \beta}\,,
\label{2.3}\\
&&
{\cal L}^{(4,4)}_{\Pom \Pom f_{1}} = \frac{g''_{\Pom \Pom f_{1}}}{24 \cdot 32 \cdot M_{0}^{4}}
\Big( \Pom_{\kappa \lambda}
\twosidep{\mu_{1}} \twosidep{\mu_{2}} \twosidep{\mu_{3}} \twosidep{\mu_{4}}
\Pom_{\rho \sigma} \Big)
\Big( \p_{\alpha} U_{\beta} - \p_{\beta} U_{\alpha} \Big)\,
\Gamma^{(10)\,\kappa \lambda, \rho \sigma, \mu_{1} \mu_{2} \mu_{3} \mu_{4}, \alpha \beta}\,, \;\;\quad
\label{2.4}
\end{eqnarray}
where $M_{0} \equiv 1$~GeV (introduced for dimensional reasons),
$g'_{\Pom \Pom f_{1}}$ and $g''_{\Pom \Pom f_{1}}$ are
dimensionless coupling constants, 
$\Pom_{\kappa \lambda}$ is the $\Pom$ effective field,
$U_{\alpha}$ is the $f_{1}$ field,
and $\Gamma^{(8)}$, $\Gamma^{(10)}$ 
are known tensor functions~\cite{Lebiedowicz:2020yre}.

\textbf{(2)} Our second approach uses holographic QCD, in particular
the Sakai-Sugimoto model \cite{Sakai:2004cn,Brunner:2015oqa,Anderson:2014jia}
where the $\Pom \Pom f_{1}$ coupling is determined 
by the mixed axial-gravitational anomaly of QCD.
In this approach (see Appendix~B of \cite{Lebiedowicz:2020yre})
\begin{eqnarray}
{\cal L}^{\rm CS} = \varkappa' \,U_{\alpha}\,\varepsilon^{\alpha \beta \gamma \delta}\,
\Pom^{\mu}_{\;\;\beta}\, \p_{\delta}\Pom_{\gamma \mu}
+ \varkappa'' \,U_{\alpha}\,\varepsilon^{\alpha \beta \gamma \delta}\,
\left( \p_{\nu}\Pom^{\mu}_{\;\;\beta} \right) 
\left( \p_{\delta}\p_{\mu}\Pom^{\nu}_{\;\;\gamma} - \p_{\delta}\p^{\nu}\Pom_{\gamma \mu} \right)
\label{2.5}
\end{eqnarray}
with $\varkappa'$ a dimensionless constant and
$\varkappa''$ a constant of dimension GeV$^{-2}$.
For the CEP reaction,
we use the $\Pom \Pom f_{1}$ vertex derived from (\ref{2.5})
supplemented by suitable form factor (\ref{Fpompommeson_exp}). 

For our fictitious reaction ($\Pom + \Pom \to f_{1}$) there is
strict equivalence
${\cal L}^{{\rm CS}} \wideestimates 
{\cal L}^{(2,2)} + {\cal L}^{(4,4)}$
if the couplings satisfy the relations
\begin{eqnarray}
g'_{\Pom \Pom f_{1}} =
-\varkappa'\,\frac{M_{0}^{2}}{k^{2}}
-\varkappa''\,\frac{M_{0}^{2}(k^{2}-2 m^{2})}{2k^{2}} \,,
\qquad
g''_{\Pom \Pom f_{1}} =
\varkappa''\,\frac{2 M_{0}^{4}}{k^{2}} \,.
\label{2.7}
\end{eqnarray}
For our CEP reaction (\ref{1.1}) we are dealing with pomerons of mass
squared $t_{1}, t_{2} < 0$ and, in general, $t_{1} \neq t_{2}$.
Then, the equivalence relation
for small values $|t_{1}|$ and $|t_{2}|$
will still be approximately true and we confirm this
by explicit numerical studies 
(see Fig.~11 of \cite{Lebiedowicz:2020yre}).

For realistic applications we should multiply the ``bare'' vertex
$\Gamma^{(\Pom \Pom f_{1})}(q_{1},q_{2})$
as derived from a corresponding coupling Lagrangian
by a form factor $\tilde{F}^{(\Pom \Pom f_{1})}(t_{1},t_{2},k^{2})$
which we take in the factorised ansatz
\begin{eqnarray}
\tilde{F}^{(\Pom \Pom f_{1})}(t_{1},t_{2},m_{f_{1}}^{2}) = 
\exp\left( \frac{t_{1}+t_{2}}{\Lambda_{E}^{2}}\right) \,,
\label{Fpompommeson_exp}
\end{eqnarray}
where the cutoff constant $\Lambda_{E}$ 
should be adjusted to experimental data.

As discussed in Appendix~B of \cite{Lebiedowicz:2020yre},
the prediction for $\varkappa''/\varkappa'$
obtained in the Sakai-Sugimoto model is
\begin{equation}
\varkappa''/ \varkappa' = -(6.25 \cdots 2.44) \;\mathrm{GeV}^{-2}\label{kapparatiorange}
\end{equation}
for $M_\mathrm{KK}=(949 \cdots 1532)\;\mathrm{MeV}$.
Usually \cite{Sakai:2004cn} $M_{\text{KK}}$ 
is fixed by matching the mass of the lowest vector meson 
to that of the physical $\rho$ meson,
leading to $M_{\text{KK}}=949\,\text{MeV}$.
However, this choice leads to a tensor glueball mass 
which is too low, $M_T \approx 1.5\,\text{GeV}$.
The pomeron trajectory 
[$\alpha_{\Pom}(t) = \alpha_{\Pom}(0)+\alpha'_{\Pom} t$, 
$\alpha_{\Pom}(0) = 1.0808$,
$\alpha'_{\Pom} = 0.25$~GeV$^{-2}$]
corresponds to $M_T \approx 1.9\,\text{GeV}$,
whereas lattice predictions correspond to $M_T \gtrsim 2.4\,\text{GeV}$.

\section{Results}
\label{sec:results}

\subsection{Comparison with the WA102 data}

\begin{figure}[!ht]
\centering
\includegraphics[width=0.4\textwidth]{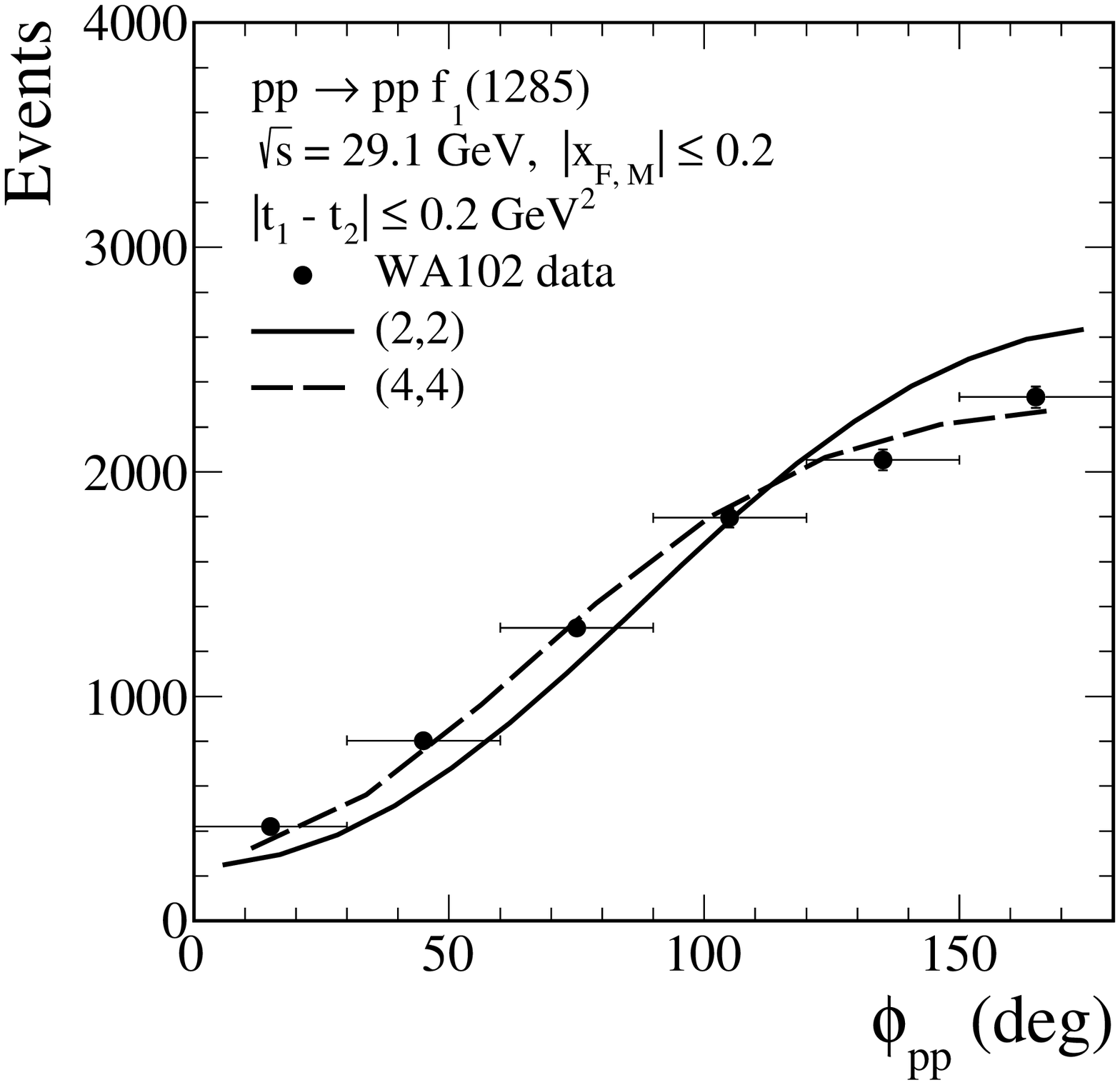}
\includegraphics[width=0.4\textwidth]{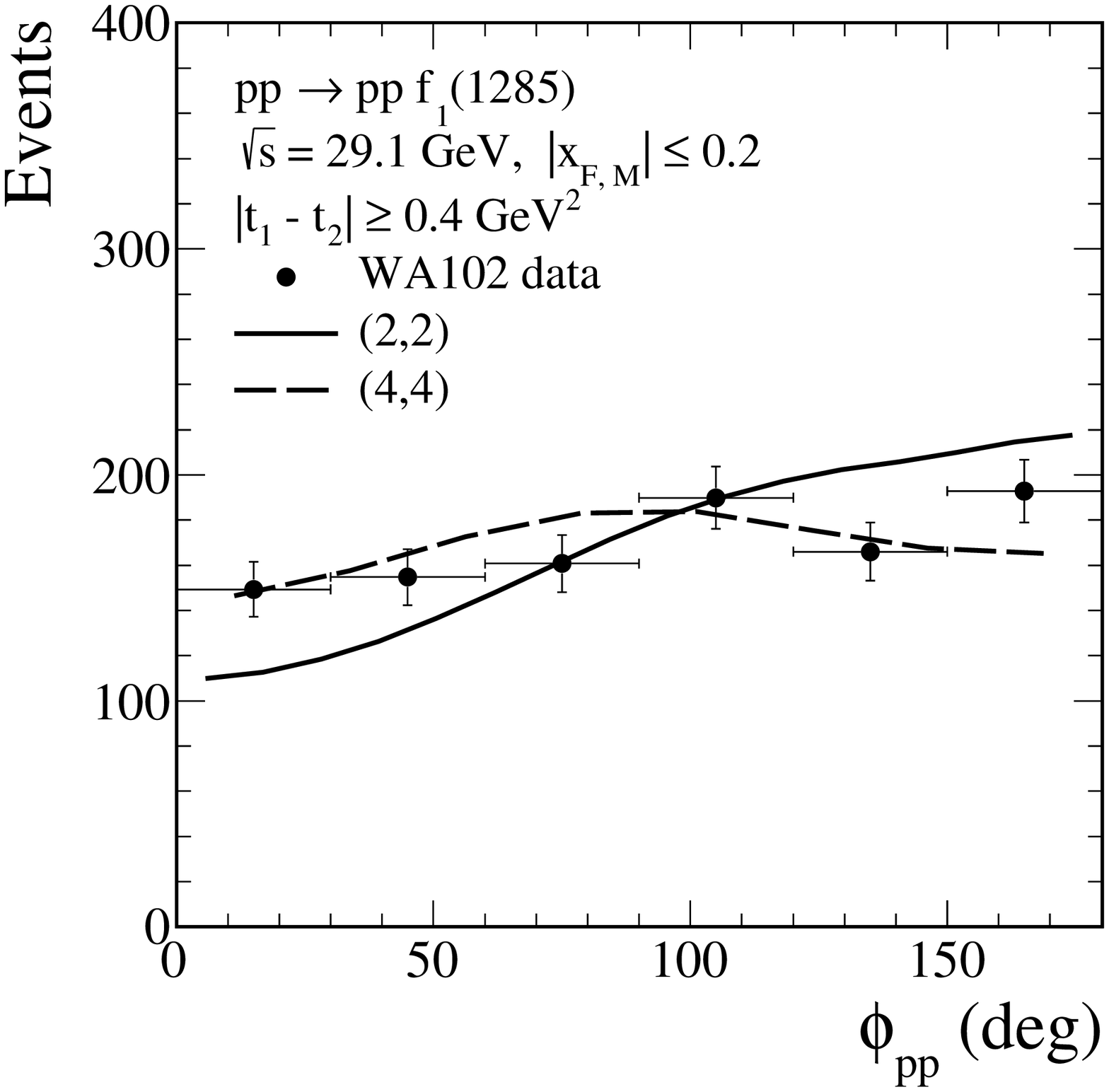}\\
\includegraphics[width=0.4\textwidth]{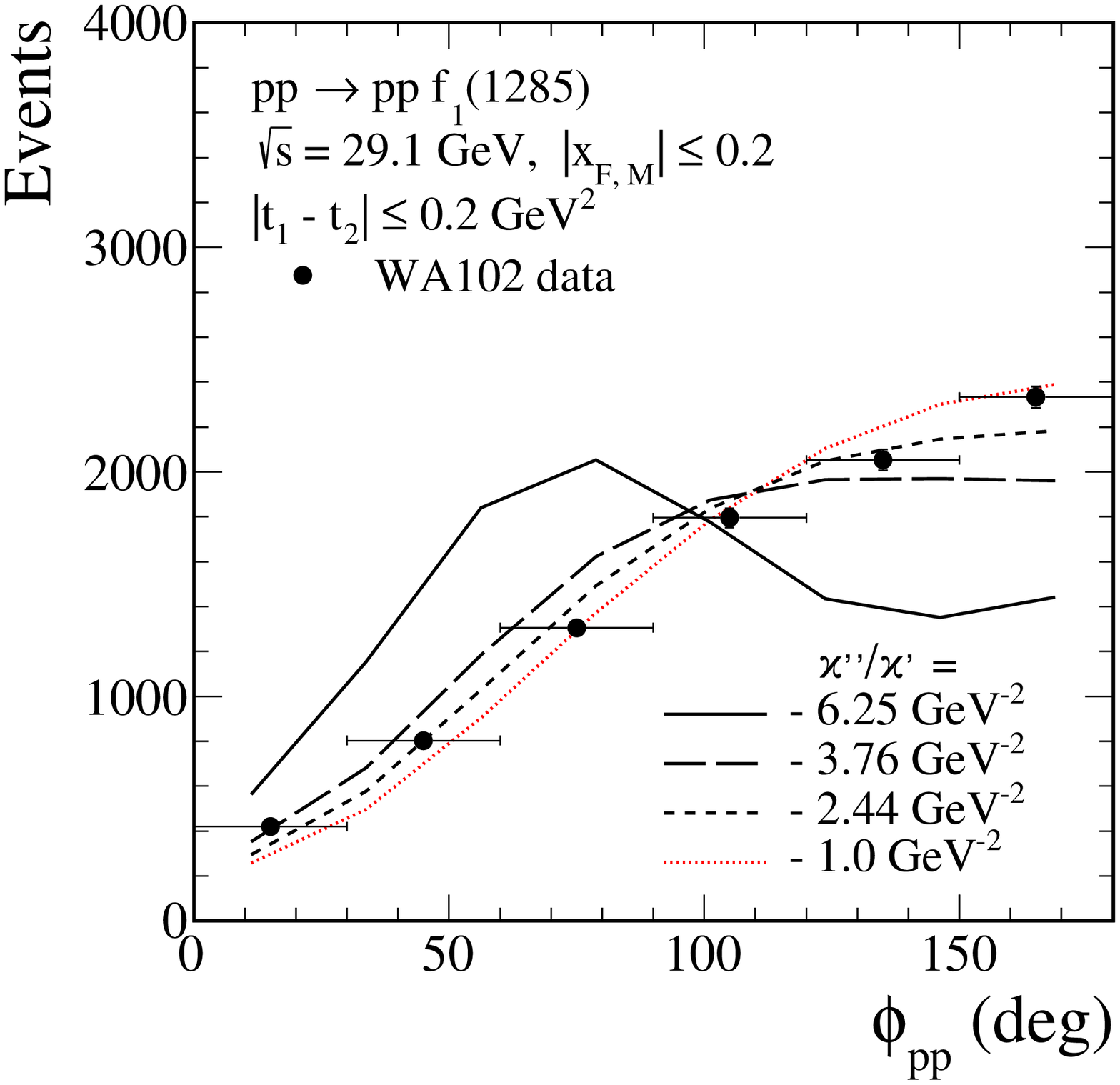}
\includegraphics[width=0.4\textwidth]{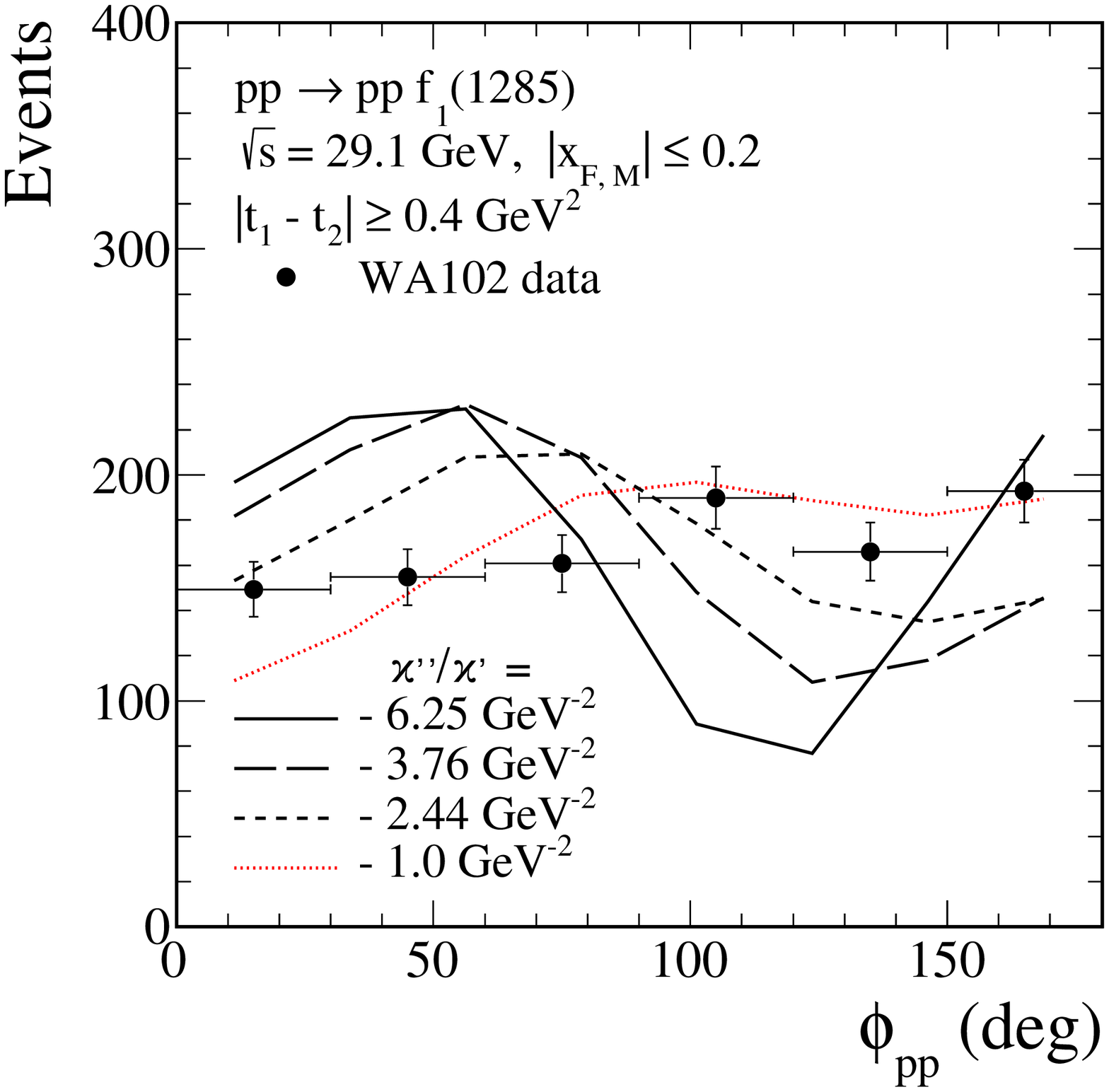}
\caption{\label{fig2}
\small
The $\phi_{pp}$ distributions for $f_{1}(1285)$ meson production
at $\sqrt{s} = 29.1$~GeV, $|x_{F,M}| \leqslant 0.2$, and
for $|t_{1} - t_{2}| \leqslant 0.2$~GeV$^{2}$ (left panels)
and $|t_{1} - t_{2}| \geqslant 0.4$~GeV$^{2}$ (right panels).
The WA102 experimental data points are from Fig.~3 of \cite{Kirk:1999df}.
The theoretical results
have been normalised to the mean value of the number of events.
The results for $\Lambda_{E} = 0.7$~GeV a form-factor parameter (\ref{Fpompommeson_exp}) are shown.}
\end{figure}

The WA102 collaboration obtained 
for the $pp \to pp f_{1}(1285)$ reaction 
the total cross section of
$\sigma_{\rm exp} = (6919 \pm 886)\;\mathrm{nb}$
at $\sqrt{s} = 29.1$~GeV and
for a cut on the central system $|x_{F}| \leqslant 0.2$ 
\cite{Barberis:1998by}.
The WA102 collaboration also gave distributions in $t$ and
in $\phi_{pp}$ ($0 \leqslant \phi_{pp} \leqslant \pi$),
the azimuthal angle between the transverse momenta
of the two outgoing protons.
We are assuming that the reaction (\ref{1.1})
is dominated by pomeron exchange already
at $\sqrt{s} = 29.1$~GeV.
In \cite{Kirk:1999df} an interesting behaviour 
of the $\phi_{pp}$ distribution for $f_{1}(1285)$ meson production 
for two different values of $|t_{1} - t_{2}|$ was presented.
In Figure~\ref{fig2} we show 
some of our results \cite{Lebiedowicz:2020yre}
which include absorptive corrections; 
see Eqs.~(\ref{amp_full}), (\ref{abs_correction}).
We show the $\phi_{pp}$ distribution of events
from \cite{Kirk:1999df}
for $|t_{1} - t_{2}| \leqslant 0.2$~GeV$^{2}$ (left panels) and 
$|t_{1} - t_{2}| \geqslant 0.4$~GeV$^{2}$ (right panels).
From the top panels, it seems that the $(l,S) = (4,4)$ term (\ref{2.4})
best reproduces the shape of the WA102 data.
The absorption effects play a significant role there.
In the bottom panels of Fig.~\ref{fig2} 
we examine the combination of two 
$\Pom \Pom f_{1}$ couplings $\varkappa'$ and $\varkappa''$
calculated with the vertex (\ref{2.5}).
The ratio (\ref{kapparatiorange}) agrees with the fit 
$\varkappa''/\varkappa' = -1.0$~GeV$^{-2}$
as far as the sign of this ratio is
concerned, but not in its magnitude.
This could indicate that the Sakai-Sugimoto model needs a more complicated
form of reggeization of the tensor glueball propagator
as indeed discussed in \cite{Anderson:2014jia}
in the context of CEP of $\eta$ and $\eta'$ mesons.
It could also be an indication 
of the importance of secondary contributions with reggeon exchanges,
i.e. $\Reg \Reg$-, $\Reg \Pom$-, and $\Pom \Reg$-fusion processes.

We get a reasonable description of the WA102 data
with $\Lambda_{E} = 0.7$~GeV and the following possibilities:
\begin{eqnarray}
(l,S) = (2,2)\;\mathrm{term \; only}:&&
g'_{\Pom \Pom f_{1}} = 4.89\,, \;
g''_{\Pom \Pom f_{1}} = 0;
\label{3.2}\\
(l,S) = (4,4)\;\mathrm{term \; only}:&&
g'_{\Pom \Pom f_{1}} = 0\,, \;
g''_{\Pom \Pom f_{1}} = 10.31;
\label{3.3}\\
\mathrm{CS \; terms}:&&
\varkappa' = -8.88\,, \;
\varkappa''/\varkappa' = -1.0\;\mathrm{GeV}^{-2}\,.
\label{3.4}
\end{eqnarray}
Now we can use our equivalence relation (\ref{2.7})
in order to see to which $(l,S)$ couplings (\ref{3.4})
corresponds.
Replacing in (\ref{2.7}) $m^{2}$ by 
$t_{1} = t_{2} = -0.1$~GeV$^{2}$
and $k^{2}$ by $m_{f_{1}}^{2} = (1282 \;\mathrm{MeV})^{2}$
we get from (\ref{3.4})
\begin{equation}
g'_{\Pom \Pom f_{1}} = 0.42\,, \;
g''_{\Pom \Pom f_{1}} = 10.81\,.
\label{3.5}
\end{equation}
Thus, the CS couplings of (\ref{3.4}) correspond
to a nearly pure $(l,S) = (4,4)$ coupling (\ref{3.3}).

\begin{figure}[!ht]
\centering
\includegraphics[width=0.38\textwidth]{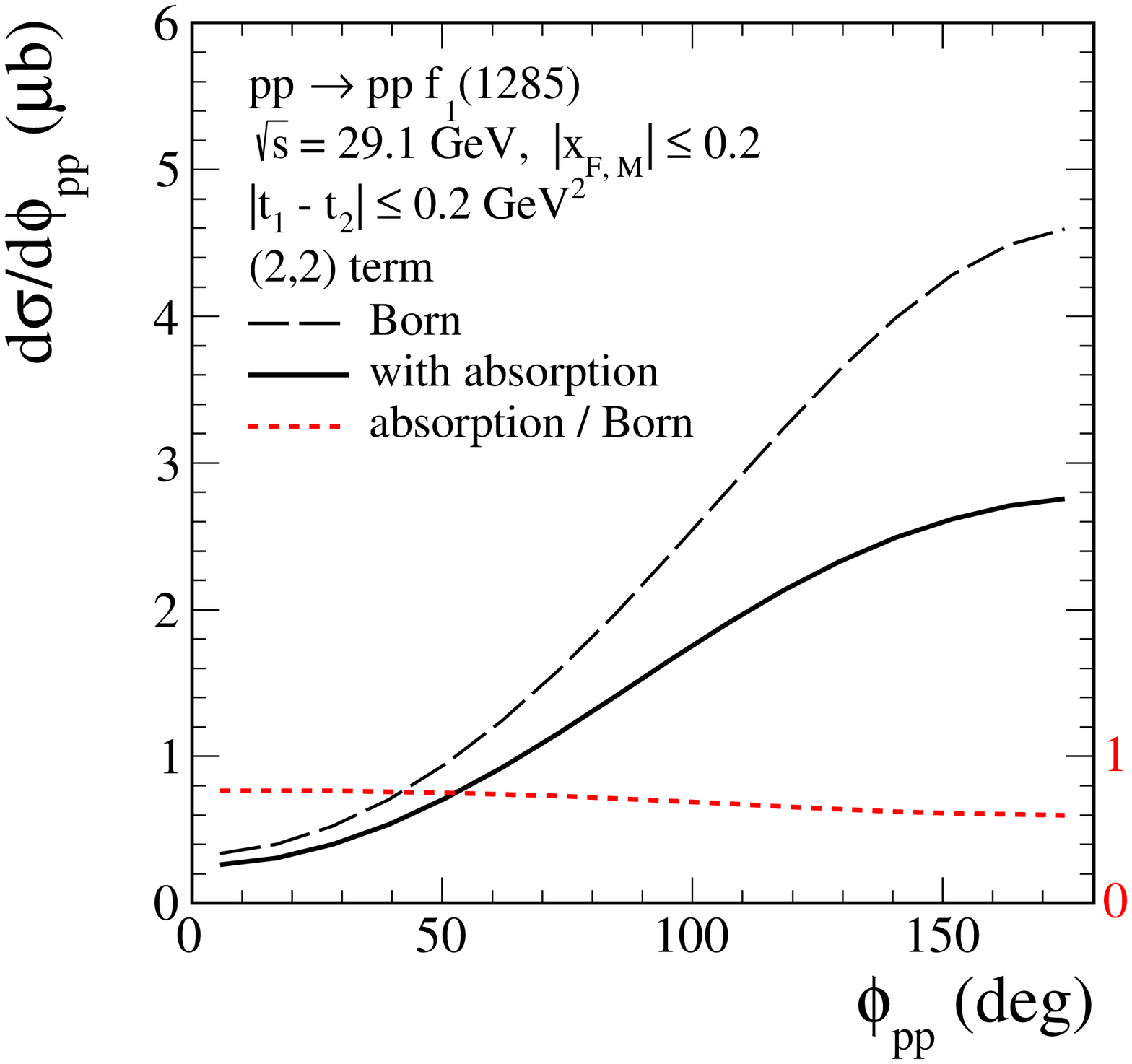}
\includegraphics[width=0.38\textwidth]{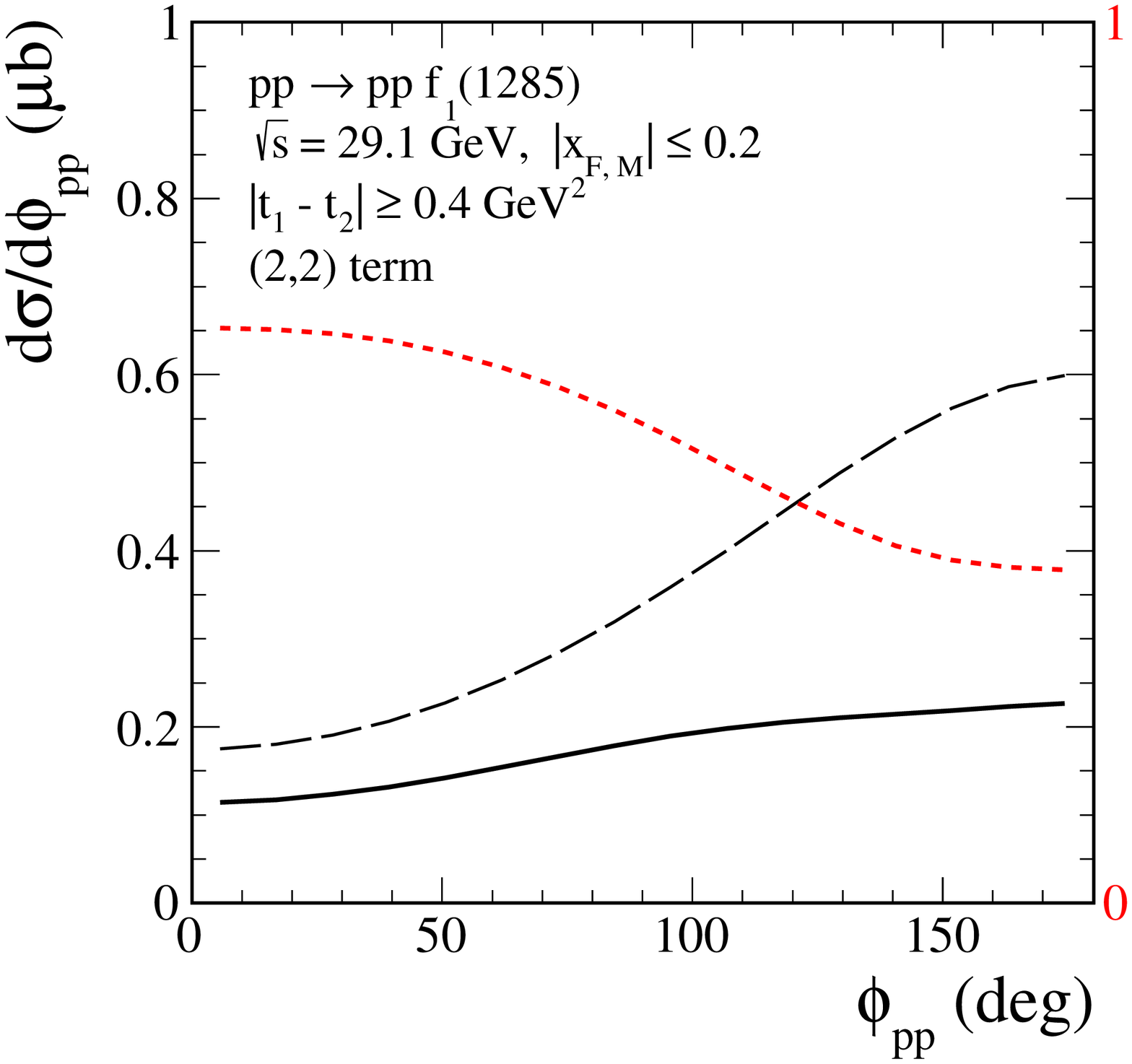}
\includegraphics[width=0.38\textwidth]{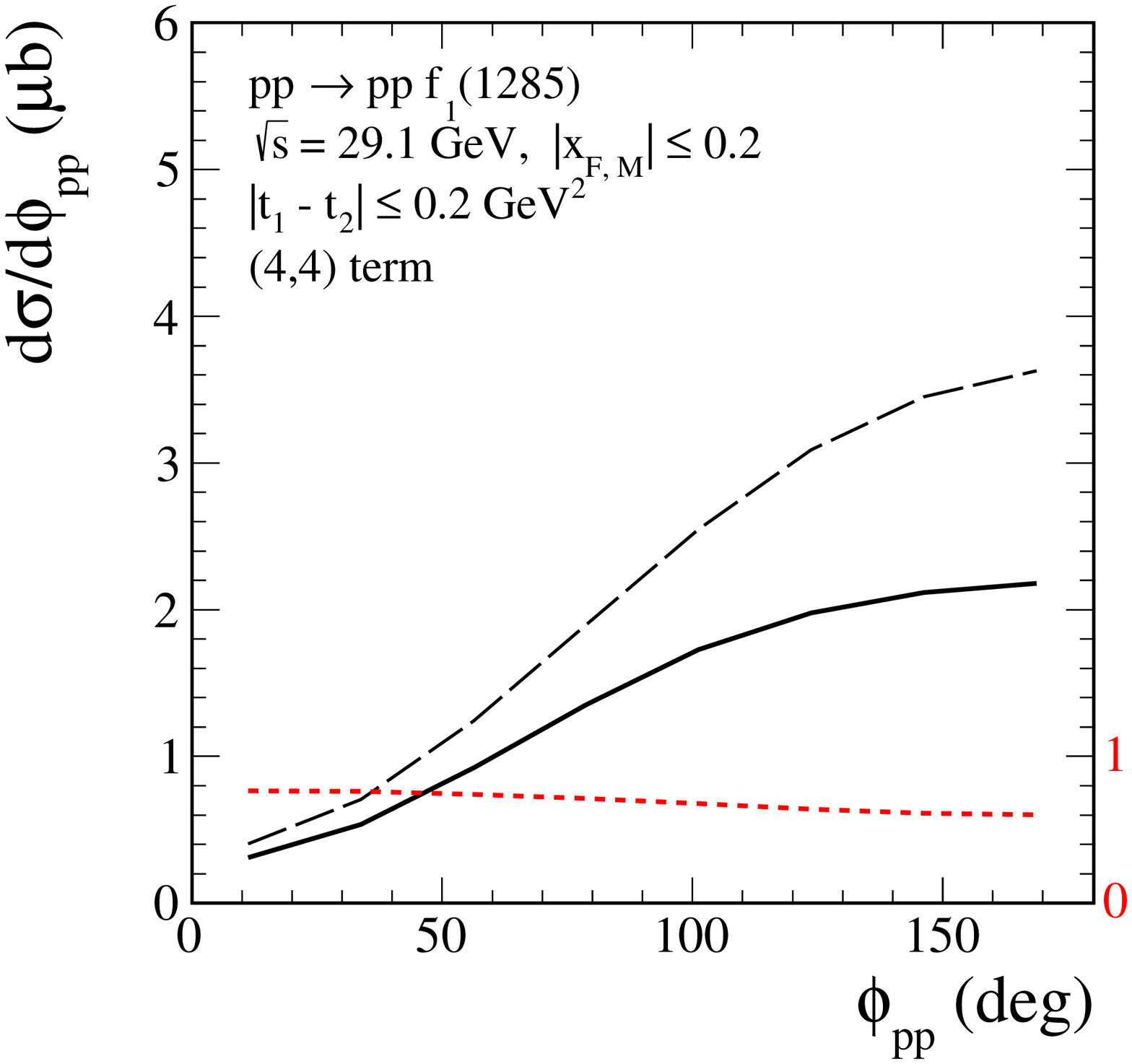}
\includegraphics[width=0.38\textwidth]{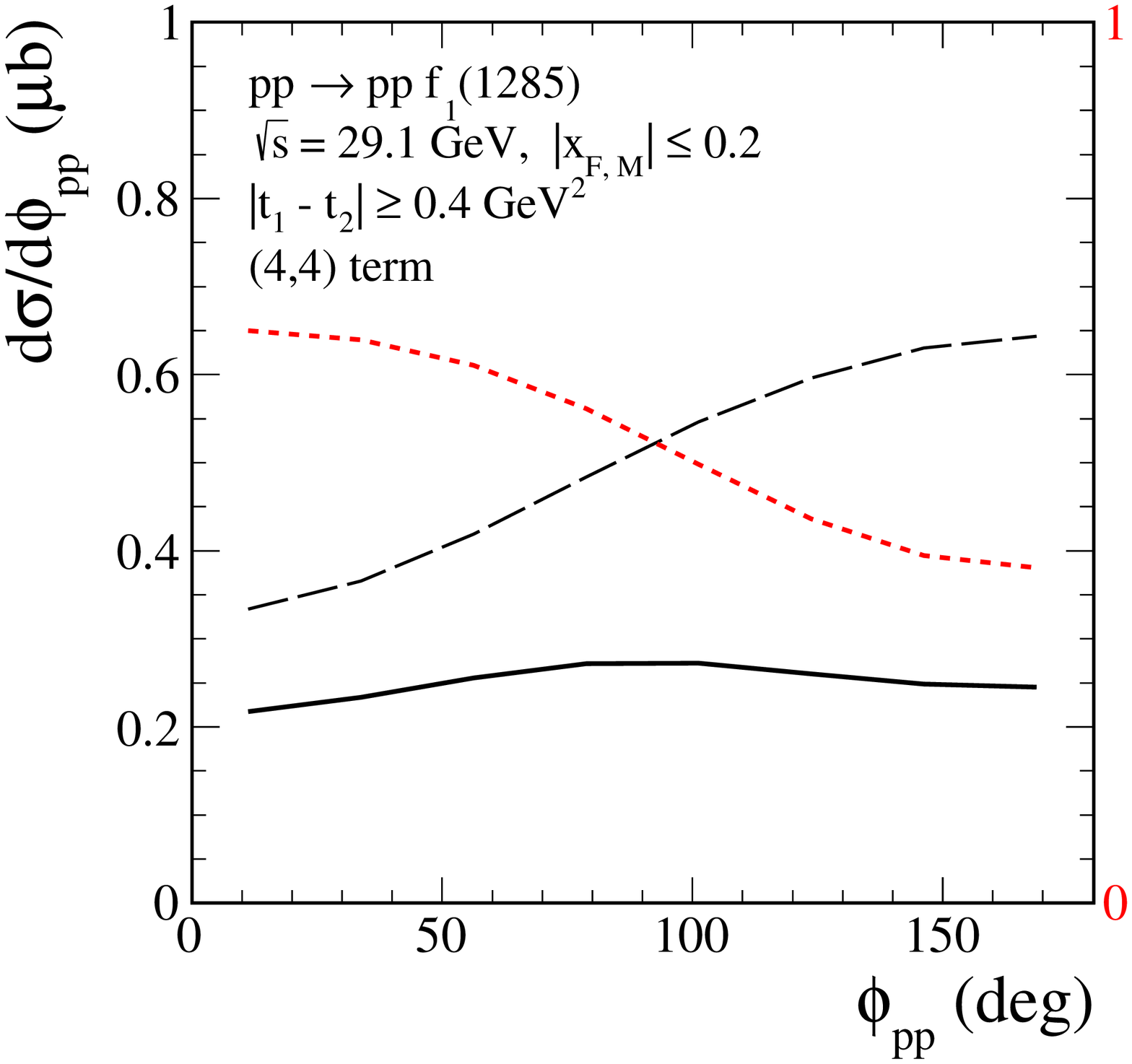}
\caption{\label{2abs}
\small
The $\phi_{pp}$ distributions for $f_{1}(1285)$ meson production
at $\sqrt{s} = 29.1$~GeV, $|x_{F,M}| \leqslant 0.2$,
for $|t_{1} - t_{2}| \leqslant 0.2$~GeV$^{2}$ (left) and for
$|t_{1} - t_{2}| \geqslant 0.4$~GeV$^{2}$ (right).
The long-dashed black lines represent the Born results
and the solid black lines correspond to the results with the absorption effects included.
The dotted red lines represent the ratio of full and Born cross sections on the scale indicated by the red numbers on the r.h.s. of the panels.}
\end{figure}

In Figure~\ref{2abs} we show the results for the $\phi_{pp}$ distributions
for different cuts on $|t_{1} - t_{2}|$ without and with the absorption
effects included in the calculations.
The results for the two $(l,S)$ couplings are shown.
The absorption effects lead to a large reduction of the cross section.
We obtain the ratio of full and Born cross sections,
the survival factor, as $\langle S^{2}\rangle = 0.5$--$0.7$.
Note that $\langle S^{2}\rangle$ depends on the kinematics.
We can see a large damping of the cross section 
in the region of $\phi_{pp} \sim \pi$,
especially for $|t_{1} - t_{2}| \geqslant 0.4$~GeV$^{2}$.
We notice that our results for the $(4,4)$ term 
have similar shapes as those presented in \cite{Petrov:2004hh} 
[see Figs.~3(c) and 3(d)]
where the authors also included the absorption corrections.

Having fixed the parameters of the model in this way 
we will give predictions for the LHC experiments.
Because of the possible influence of nonleading exchanges
at low energies, these predictions for cross sections 
at high energies should be regarded rather as an upper limit.
The secondary reggeon exchanges should give small contributions at high energies and in the midrapidity region.
As discussed in Appendix~D of \cite{Lebiedowicz:2020yre}
we expect that they should overestimate the cross sections
by not more than a factor of 4.

\subsection{Predictions for the LHC experiments}

\begin{figure}[h]
\centering
\includegraphics[width=0.4\textwidth]{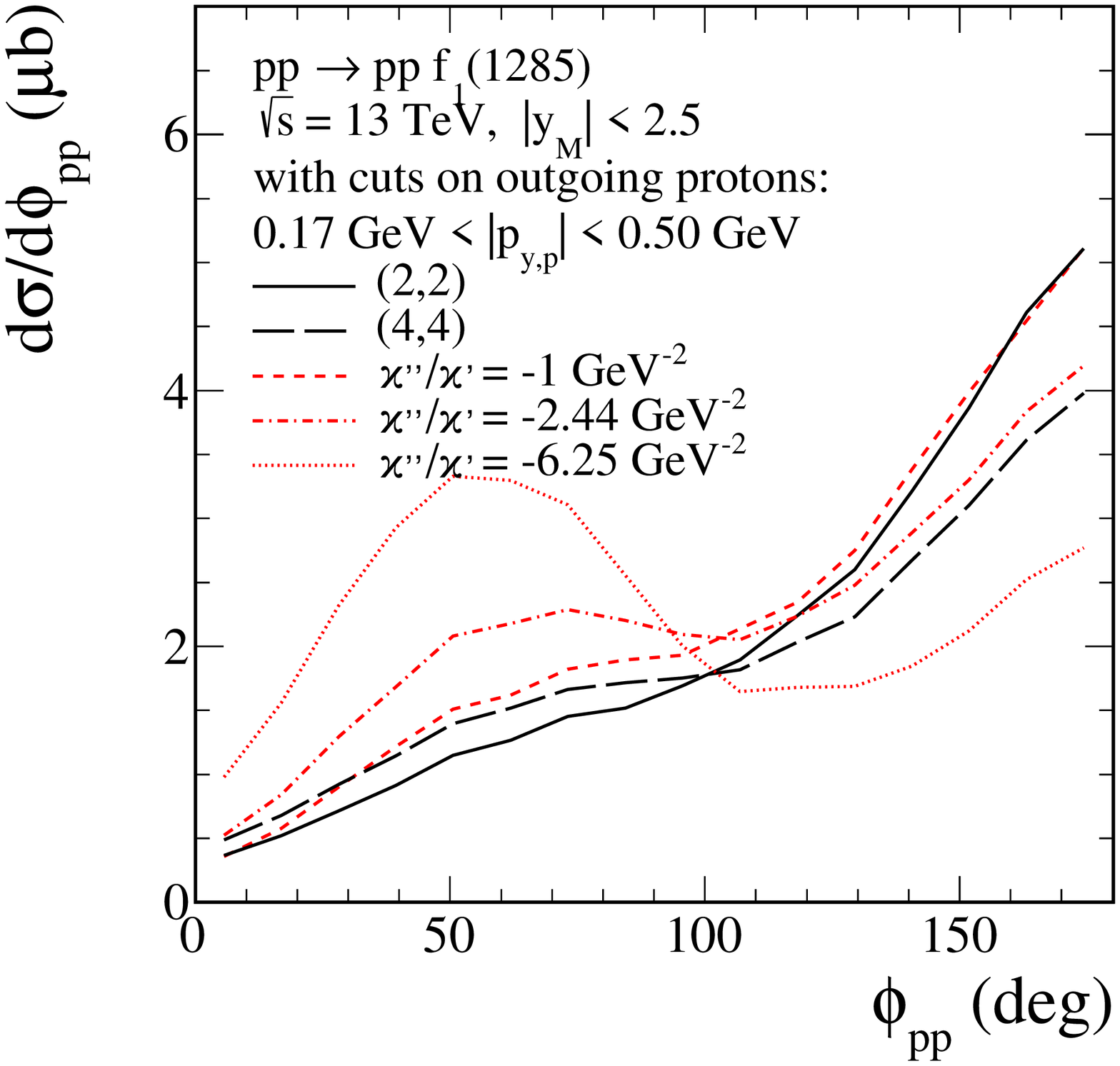}
\includegraphics[width=0.4\textwidth]{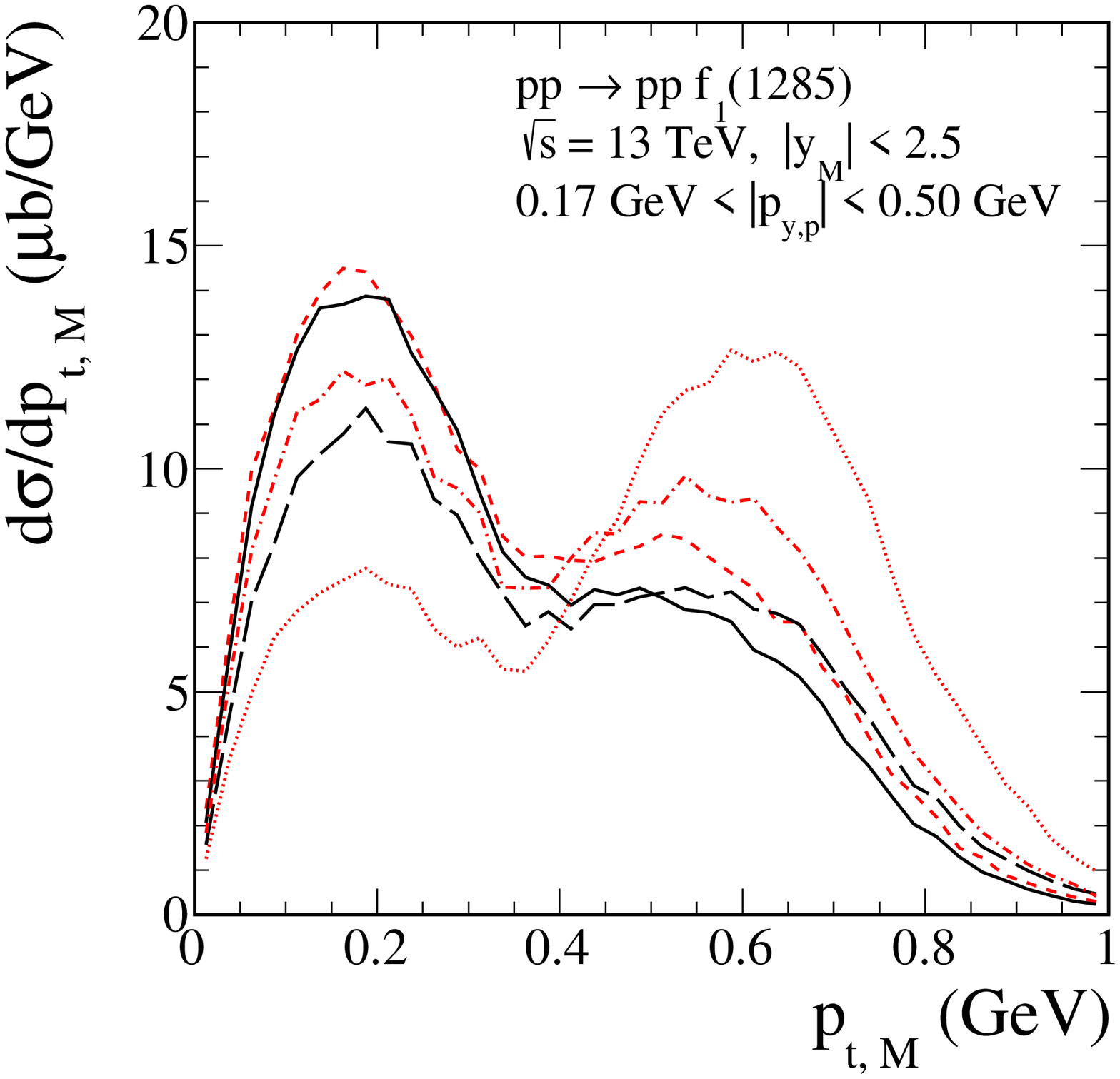}
\caption{\label{fig:ATLAS-ALFA}
The differential cross sections for
the $f_{1}(1285)$ production at $\sqrt{s} = 13$~TeV, 
$|{\rm y_{M}}| < 2.5$,
and with cuts on momenta of outgoing protons
($0.17\;{\rm GeV} < |p_{y,p}| < 0.50\;{\rm GeV}$).
The results for $(l,S)$ and $(\varkappa',\varkappa'')$ terms are shown.}
\end{figure}

Now we wish to show (selected) results 
for the $pp \to pp f_{1}(1285)$ reaction for the LHC;
see \cite{Lebiedowicz:2020yre} for many more results.
In Figure~\ref{fig:ATLAS-ALFA} we show our predictions 
for the distributions of $\phi_{pp}$
and the transverse momentum of the $f_{1}(1285)$
for $\sqrt{s} = 13$~TeV, $|{\rm y_{M}}| < 2.5$,
and for the cut on the leading protons of
$0.17\;{\rm GeV} < |p_{y,p}| < 0.50\;{\rm GeV}$.
The results for the $(l,S) = (2,2)$ term (\ref{2.3}),
the $(4,4)$ term (\ref{2.4}),
and for the $\varkappa'$ plus $\varkappa''$ terms 
calculated with (\ref{2.5}) for (\ref{kapparatiorange})
obtained in the Sakai-Sugimoto model 
(see Appendix~B of \cite{Lebiedowicz:2020yre}) are shown.
For comparison, the results for our fit to WA102 data
($\varkappa''/\varkappa'=-1.0$~GeV$^{-2}$) are also presented.
The contribution with $\varkappa''/\varkappa' = -6.25$~GeV$^{2}$
gives a significantly different shape.
This could be tested in experiments, 
such as ATLAS-ALFA \cite{Sikora:2020mae},
when both protons are measured.
We obtain the ratio of full and Born cross sections
as $\langle S^{2}\rangle \simeq 0.3$ for $\sqrt{s} = 13$~TeV.

The four-pion decay channel seems well suited to measure 
the CEP of the $f_{1}(1285)$ at the LHC \cite{Sikora:2020mae}.
We predict a large cross section 
for the exclusive axial-vector $f_{1}(1285) \to 4 \pi$ production 
compared to the CEP of the tensor $f_{2}(1270) \to 4 \pi$
\cite{Lebiedowicz:2016ioh,Lebiedowicz:2019por}.
The $4 \pi$ continuum for the $pp \to pp 4 \pi$
reaction was studied
in \cite{Lebiedowicz:2016zka,Kycia:2017iij}.

\section{Conclusion}
\label{sec:conclusions}

\begin{enumerate}
\item[$\bullet$] 
The calculations for the $pp \to ppf_{1}(1285)$ reaction
have been performed in the tensor-pomeron approach \cite{Ewerz:2013kda}.
We have discussed in detail the forms of the $\Pom \Pom f_{1}$
coupling. 
Detailed tests of the Sakai-Sugimoto model are possible.

\item[$\bullet$] We obtain a good description of the WA102 data at
$\sqrt{s} = 29.1$~GeV \cite{Barberis:1998by,Kirk:1999df} 
assuming that the $pp \to pp f_{1}(1285)$ reaction 
is dominated by pomeron-pomeron fusion.

\item[$\bullet$] We obtain a large cross section for CEP of
the $f_{1}(1285)$ of
$\sigma \cong 6-40 \;\mu\mathrm{b}$
for the ALICE, ATLAS-ALFA, CMS, and LHCb experiments,
depending on the assumed cuts (see Table~III of~\cite{Lebiedowicz:2020yre}).
Predictions for the STAR experiments at RHIC are also given
in \cite{Lebiedowicz:2020yre}.
In all cases the absorption effects were included.

\item[$\bullet$]
Experimental studies of single meson CEP reactions will allow
to extract many $\Pom \Pom M$ coupling parameters.
The holographic methods applied to QCD already give some predictions \cite{Anderson:2014jia,Lebiedowicz:2020yre}.

\item[$\bullet$]
Detailed analysis of the distributions in  $\phi_{pp}$,
the azimuthal angle between the transverse momenta of the outgoing protons,
can help to solve several important problems
for soft processes,
to check/study the real pattern of the interaction (absorption models),
to understand the difference in the dynamics 
of production of $q\bar{q}$ mesons and glueballs 
(or more accurately, states which are believed to have a large glueball component),
to disentangle $f_{1}$- and $\eta$-type resonances 
contributing to the same final channel.

\item[$\bullet$] 
Such studies could be extended, for instance by the COMPASS experiment
where presumably one could study the influence of
reggeon-pomeron and reggeon-reggeon fusion terms.
Future experiments available at the GSI-FAIR with HADES and PANDA
should provide new information about the $\rho \rho f_{1}$
and $\omega \omega f_{1}$ couplings \cite{Lebiedowicz:2021gub}.

\end{enumerate}

\section*{Acknowledgements}
The author is indebted to A. Szczurek, O. Nachtmann, A. Rebhan, 
and J. Leutgeb for their cooperation on this topic.
Many thanks to R. Ryutin and V. Petrov
for inviting me to the XXXIII International (ONLINE) Workshop 
on High Energy Physics and for useful discussions.




\bibliography{SciPost_Example_BiBTeX_File.bib}

\nolinenumbers

\end{document}